\documentclass[sigconf]{acmart}

\AtBeginDocument{%
  \providecommand\BibTeX{{%
    \normalfont B\kern-0.5em{\scshape i\kern-0.25em b}\kern-0.8em\TeX}}}

\setcopyright{acmcopyright}
\copyrightyear{2019}
\acmYear{2019}
\acmDOI{}

\acmConference[SIGGRAPH Asia 2019]{The 12th ACM SIGGRAPH Conference and Exhibition on Computer Graphics and Interactive Techniques in Asia}{November 17 -- 20, 2019}{Brisbane, Australia}
\acmBooktitle{The 12th ACM SIGGRAPH Conference and Exhibition on Computer Graphics and Interactive Techniques in Asia, November 17 -- 20, 2019, Brisbane, Australia}
\acmPrice{}
\acmISBN{}

\acmSubmissionID{}

\usepackage{hyperref}

\begin{document}

\title{Overt visual attention on rendered 3D objects}

\author{Oleksii Sidorov}
\email{oleksiis@stud.ntnu.no}
\orcid{0000-0002-4453-1191}
\authornote{Equal contributions.}
\affiliation{%
  \institution{The Norwegian Colour Lab}
  \institution{NTNU}
  \city{Gj{\o}vik}
  \country{Norway}
  \postcode{2815}
  \streetaddress{Teknologivegen 22}
}

\author{Joshua S. Harvey}
\authornotemark[1]
\email{joshua.harvey@pmb.ox.ac.uk}
\author{Hannah E. Smithson}
\authornotemark[1]
\email{hannah.smithson@psy.ox.ac.uk}
\affiliation{%
  \institution{Dept. of Experimental Psychology}
  \institution{University of Oxford}
  \city{Oxford}
  \country{UK}
}

\author{Jon Y. Hardeberg}
\email{jon.hardeberg@ntnu.no}
\affiliation{%
  \institution{The Norwegian Colour Lab}
  \institution{NTNU}
  \city{Gj{\o}vik}
  \country{Norway}
  \postcode{2815}
  \streetaddress{Teknologivegen 22}
}

\renewcommand{\shortauthors}{Sidorov and Harvey, et al.}

\begin{abstract}
This work covers multiple aspects of overt visual attention on 3D renders: measurement, projection, visualization, and application to studying the influence of material appearance on looking behaviour. In the scope of this work, we ran an eye-tracking experiment in which the observers are presented with animations of rotating 3D objects. The objects were rendered to simulate different metallic appearance, particularly smooth (glossy), rough (matte), and coated gold. The eye-tracking results illustrate how material appearance itself influences the observer's attention, while all the other parameters remain unchanged. In order to make visualization of the attention maps more natural and also make the analysis more accurate, we develop a novel technique of projection of gaze fixations on the 3D surface of the figure itself, instead of the conventional 2D plane of the screen. The proposed methodology will be useful for further studies of attention and saliency in the computer graphics domain.
\end{abstract}



\begin{teaserfigure}
  \includegraphics[width=\textwidth]{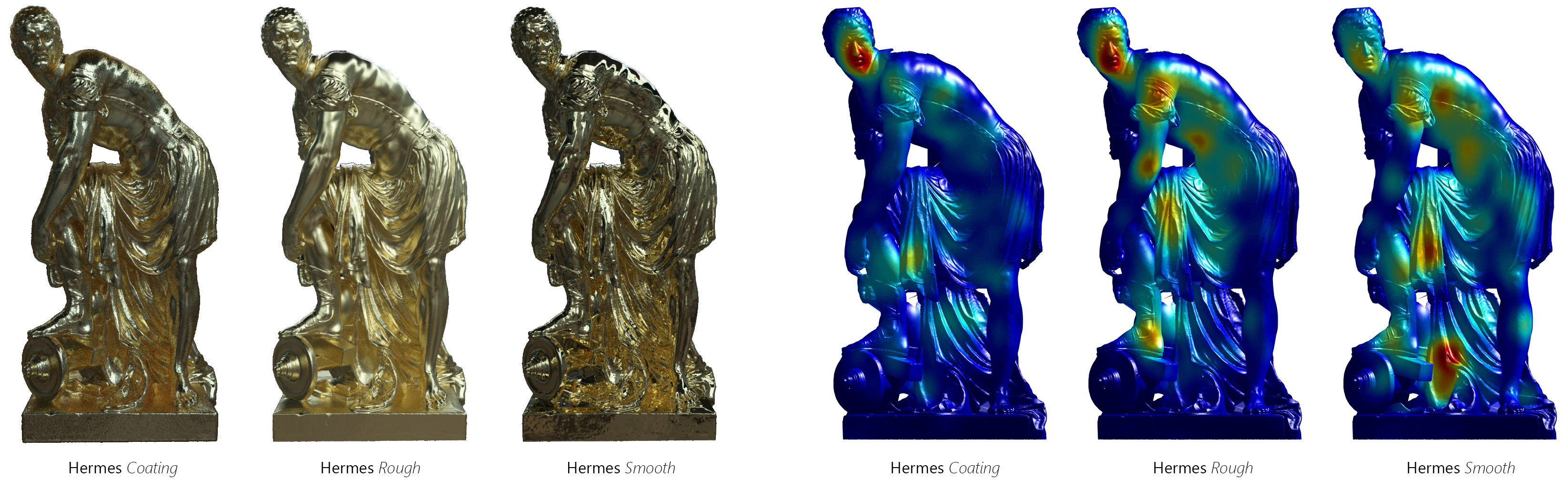}
  \caption{The rendered stimuli differing only in material appearance and corresponding fixation maps reprojected on a mesh.}
  \Description{The rendered stimuli differing only in material appearance and corresponding fixation maps reprojected on a mesh.}
  \label{fig:teaser}
\end{teaserfigure}

\maketitle

\section{Introduction}

Research on human attention and visual saliency has gained wide popularity due to the numerous tasks where it can be applied and produce useful results \cite{Jacob2003-JACETI}\cite{bergstrom2014eye}\cite{rayner2001integrating}\cite{cowen2002eye}. Such studies are particularly important for multimedia and graphic design. Measurement and prediction of the areas where the observer is likely to look helps to create content oriented specifically to the human visual system and allows low-level cognitive processes to be considered, resulting in more inclusive and efficient design. In addition, the study of saliency in computer graphics (CG) is important for understanding fundamental psychological aspects of human-computer interaction. Although the gap between rendered and natural scenes is being reduced very rapidly, in most cases, visual perception of computer graphics is still significantly different from that of natural scenes. This inevitably influences the distribution of attention which in turn complicates the transfer of well-known psychological principles from real-world to synthetic environments.\\
In this work, we develop a new methodology of more efficient projection and visualization of gaze-fixations on rendered 3D meshes and demonstrate how it can be applied on the example of material appearance. The classical eye-tracking work of Yarbus \cite{yarbus1967eye} illustrates how gaze direction is affected by the task given to the observer (or question asked), so we also perform the experiment under a set of different tasks, and analyze their impact on overt visual attention. \\
The paper is organized in the following way: Section 2 presents existing works on the study of attention in 3D as well as studies on material perception, Section 3 describes the theory behind the newly developed methodology, Section 4 presents the experimental setup and rendering parameters, and Section 5 reports eye-tracking results from the experiment and their analysis.

\section{Related works}
The novelty of our work is emphasized by the fact that there are no previous works in which authors study the same questions we try to answer. Very few recent works are dedicated to overt visual attention on 3D objects. \cite{10.1167/12.1.7} conducted a study of eye movement patterns on 3D meshes of abstract shapes in the context of studying shape perception and the influence of different features such as edges, curvatures, and concavities. Contrasting with our work, the stimuli used in that study were simple geometric shapes rendered in MatLab, and eye-tracking results were analyzed in 2D. Recent work of \cite{wang2018tracking} was aimed at assessing visual attention on the surface of real 3D objects, not CG renders. Moreover, the goal of the work was to create a large public dataset of gaze-fixations on 3D objects and to train predictive machine-learning algorithms (the resulting saliency maps were two-dimensional). Similarly to our approach, these authors project fixation maps directly on the surface of the 3D object. In the case of real objects, this process transforms into a classical task of 3D reconstruction which can be solved accurately when the geometry of the experiment is known and the system is calibrated (that is the case in \cite{wang2018tracking}). However, in the case of CG, when objects are "beyond" the screen and geometry cannot be measured such an approach cannot be used.\\ 
Several other eye-tracking studies on 3D renders have been conducted: \cite{howlett2005predicting} measured gaze fixations on low-resolution models (5000-8000 faces), \cite{kim2010mesh} evaluated the performance of 2D saliency prediction algorithms on high-res 3D models (projected onto a plane), and \cite{mantiuk2013gaze} extended the work to include experimental stimuli that were moving 3D animations. The closest match with our method can be found in work of \cite{lavoue2018visual}. The goal of the work is to create a dataset of 3D renders for benchmarking of computational algorithms of saliency prediction. As part of this endeavour, the authors perform a comprehensive study of attention under different conditions, including static and dynamic scenes. They also visualize saliency maps on the rendered 3D surfaces. According to the authors, in order to determine which surface point on the 3D mesh corresponds to the fixation pixel, they compute the ray emitted by the camera pinhole that passes through the pixel on the image, and then calculate the closest point of intersection with the 3D surface. However, it was not disclosed how the coordinates of the intersection point on a plane $\{\mathbf{x}, \mathbf{y}\}\in R^2$ were translated to the three-dimensional point cloud. Doing this requires the availability of a mapping function $G(\{\mathbf{x}, \mathbf{y}\})=\{x, y, z\}\in R^3$ in analytical form or a corresponding lookup table (LUT) which may substitute this function. We demonstrate specifically how one can estimate such a LUT and eventually perform projection on the mesh.\\
Existing work on material appearance is rather sparse, largely because it represents a rather novel research field. Color is not an exclusive characteristic that defines material perception; instead, objects may also be characterized in terms of glossiness, translucency, transparency, roughness, bumpiness, \textit{etc}. A number of works have been directed towards studying the low-level visual features that support material perception \cite{anderson2011visual}\cite{landy2007visual}\cite{fleming2014visual}. Moreover, in the majority of the cases, computer graphics is used as a tool to simulate gradation of particular appearance parameters such as glossiness or transparency. Other works have also investigated the neural processes involved in material perception. In particular, the cortical regions responsible for the perception of glossiness were studied independently by \cite{kentridge2012glossiness} and \cite{wada2014human}. A comprehensive review of advances in glossiness perception can be found in \cite{chadwick2015perception}. \cite{Toscani2013lightness} have shown a causal link between gaze behaviour and lightness judgements, and that fixations on intense pixels in rendered objects are particularly informative about surface reflectance. However, to the best of our knowledge, there are no studies of visual saliency in relation to object's appearance and glossiness in particular. \cite{qi2018fusing} collect fixation data from different materials with the aim of predicting saliency maps followed by material classification in computer vision, while \cite{leonards2007mediaeval} demonstrate how materials of different appearance are used in cultural heritage to direct the observers' attention. These works are also good examples of applications of saliency studies in computer graphics -- they are crucial for good quality simulation of particular materials, as well as for controlling observers' gaze and emphasizing certain regions in a rendered scene.

\section{Projection and visualization of fixations maps}
Let us consider a 3D-scene $S^{(3)}\in R^3$ consisting of the points $p=\{x,y,z\} \in S^{(3)}$ . Rendering the scene projects all the points of $S^{(3)}$ to the image plane: $F(S^{(3)})=S^{(2)}$ , $F(\{x,y,z\})=\{\mathbf{x},\mathbf{y}\} \in R^2$. The eye-tracker produces output as a set of fixation points in the 2D plane of a screen $\Phi\colon\varphi=\{\mathbf{x_\varphi}, \mathbf{y_\varphi}\}\in R^2$ which correspond to fixations on pixels of a rendered scene with the same location (under the condition that eye-tracker is calibrated). It may also be interpreted as the intersection point of a ray from a camera pinhole with an image plane. Therefore, the task is to find a function $G$ which re-projects fixation coordinates to a 3D space of the initial scene $G(\{\mathbf{x_\varphi}, \mathbf{y_\varphi}\})=\{x_\varphi, y_\varphi,z_\varphi\}\in S^{(3)}$ and results in 3D cloud of fixations wrapped around the scene's mesh. \\
If the projective function $F$ is known in the analytical form, the task simplifies to a trivial task of finding an inverted function $G=F^{-1}$ (which, however, may not exist). In all other cases, $G$ should be estimated. Considering that the number of the points is finite, $G$ may also be replaced with a lookup table without losing any information. Next, we propose one of the possible solutions of this task by computing a lookup table using three-dimensional \textit{color-encoding} of the coordinates. \\
The approach is based on the dimensionality of RGB color space, which is equal to the dimensionality of Euclidean space, and that therefore allows translating coordinates between them. The limitation is that RGB color space is finite (0-255 or 0-65535 depending on bit-depth), unlike Euclidean space. However, rendered 3D-scenes also occupy a strictly defined volume, which allows them to be scaled linearly to fit the limits of RGB space:

\begin{figure}[tb]
  \centering
  \includegraphics[width=\linewidth]{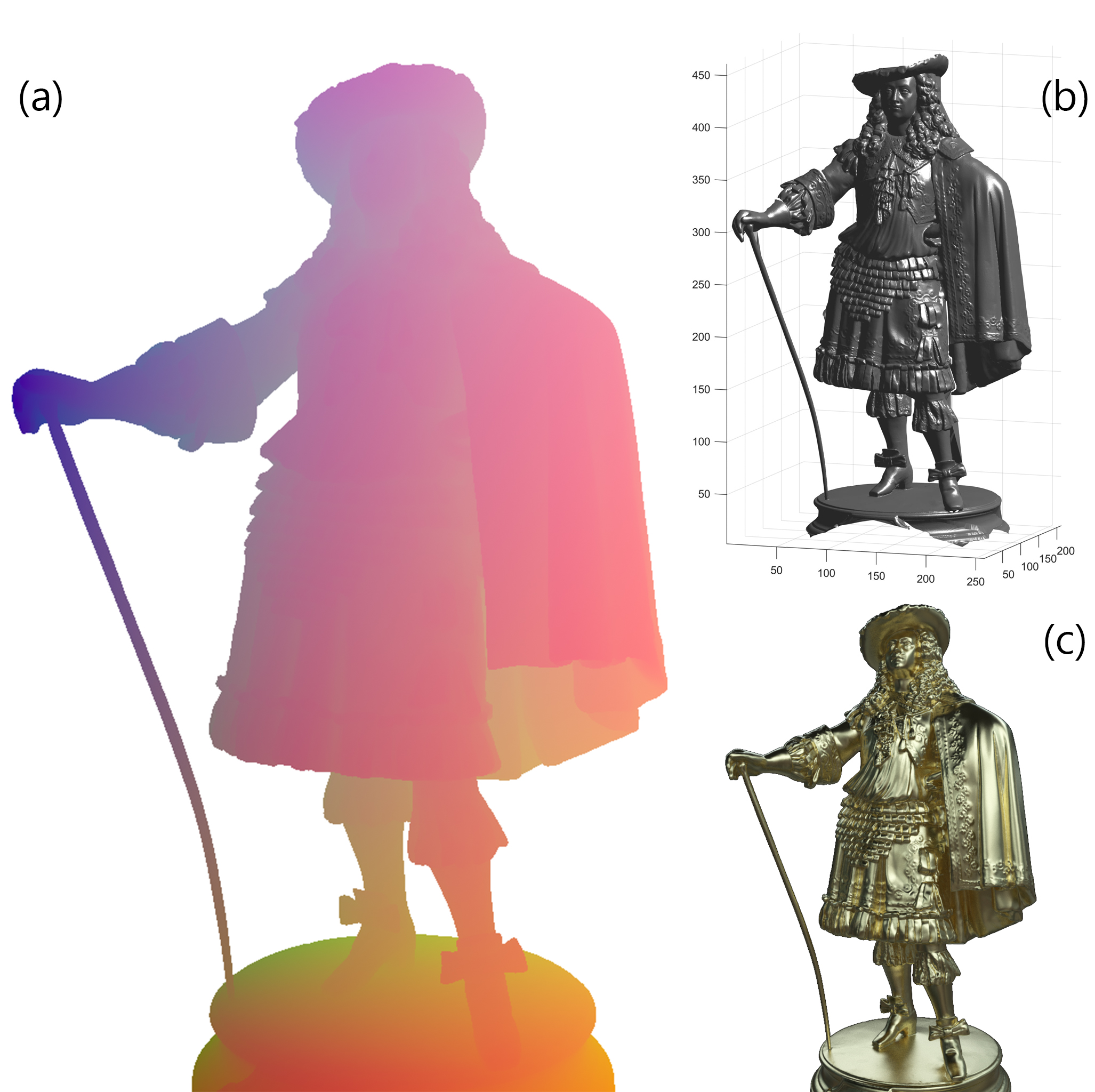}
  \caption{The proposed \textit{color encoding} of the Euclidean coordinates. (a) True-color visualization, linear RGB values are directly proportional to XYZ coordinates of each pixel (gamma-correction was applied); (b) reconstruction using image (a); (c) ground truth rendered shape.}
  \Description{The proposed color encoding of the Euclidean coordinates.}
  \label{fig:2}
\end{figure}

\begin{equation}
\begin{tabular}{c}
$R_{lin} = \frac{2^{16}\cdot(x-x_{min})}{x_{max}-x_{min}} , \quad G_{lin} = \frac{2^{16}\cdot(y-y_{min})}{y_{max}-y_{min}}, $\\ \\
$B_{lin} = \frac{2^{16}\cdot(z-z_{min})}{z_{max}-z_{min}},$
\end{tabular}
\end{equation}
considering that RGB space is a 16-bit cube (which allows spatial resolution of $2^{16}\times2^{16}\times2^{16}$). This operation forms the basis of the next step of rendering the additional map of the scene with a texture that contains information about 3D-coordinates as pixel color. This also can be interpreted as a 2D LUT. Therefore, a pixel's coordinates on an image $\{\mathbf{x},\mathbf{y}\}$  correspond to a unique color vector $G(\mathbf{x},\mathbf{y})=\{r,g,b\}$ which can be used to recover 3D information:
\begin{equation}
\left\{ 
\begin{tabular}{c}
$x=r\cdot(x_{max}-x_{min})\cdot 2^{-16}=r\cdot k_r $\\
$y=g\cdot(y_{max}-y_{min})\cdot 2^{-16}=g\cdot k_g $\\
$z = b\cdot(z_{max}-z_{min})\cdot 2^{-16}=b\cdot k_b $ 
\end{tabular} ,
\right.
\end{equation}

where $k_r,k_g$ and $k_b$ are scaling coefficients for the given scene stored separately (translation is omitted). In this case, $G\in R^2$ is a conventional RGB image which can also be transformed to sRGB for visualization (Fig. 2a). In our experiments, maps $G$ were created using Cycles render engine with custom "Emission" material of a surface configured according to (1). It is worth noting that only the points captured by a single camera view can be reconstructed using one image, similarly to real-world 3D reconstruction. Reconstruction of a whole 360$^{\circ}$ point cloud requires multi-view geometry. For example, in our experiments observers were shown only the front and sides of the object, so the rear side was not rendered and consequently was not reconstructed afterwards.  

\begin{figure*}[t]
  \centering
  \includegraphics[width=\linewidth]{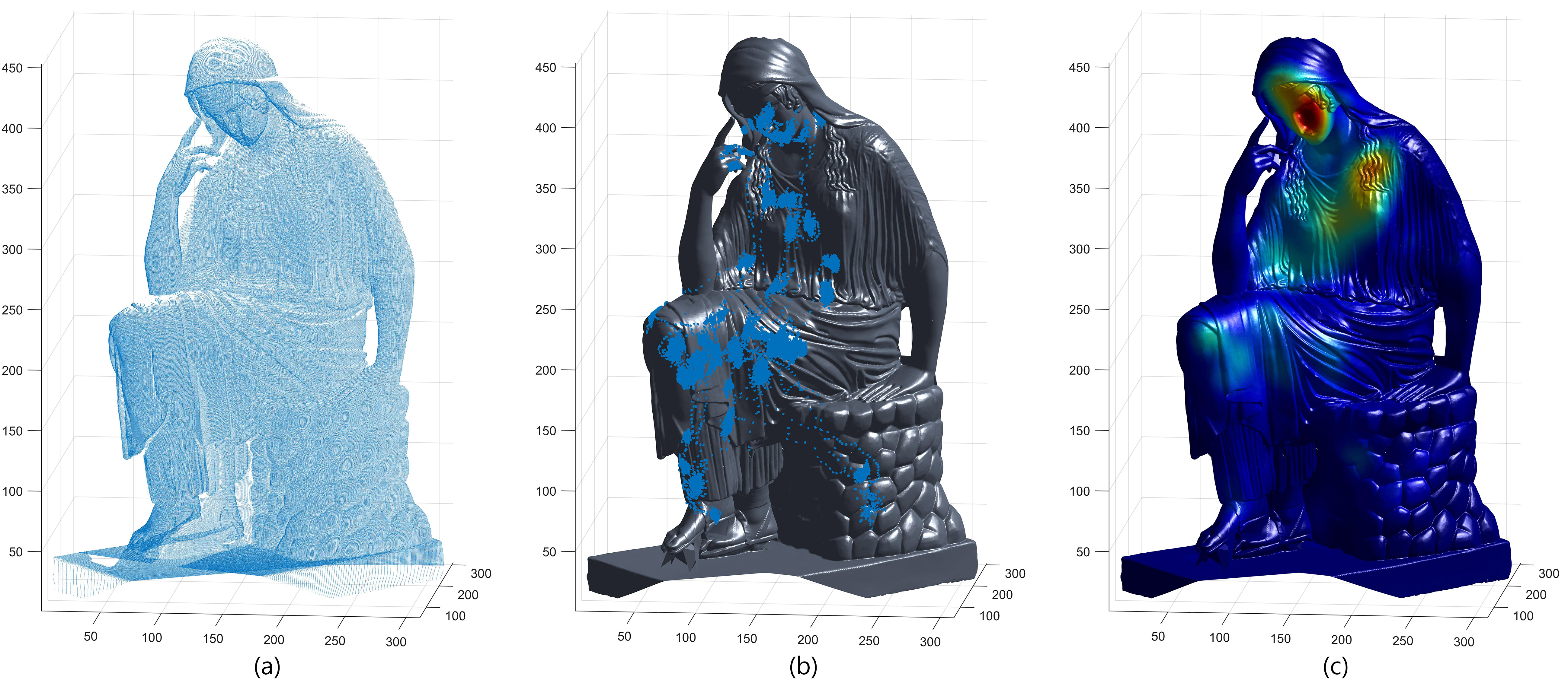}
  \caption{Surface reconstruction and re-projection of fixations via the proposed color encoding method. (a) The point cloud reconstruction; (b) gaze fixations on a mesh (only 10\% of fixations are shown); (c) output 3D saliency map.}
  \Description{Surface reconstruction and re-projection of fixations via the proposed color encoding method.}
  \label{fig:3}
\end{figure*}

\subsection{Saliency maps visualization}
The initial stage of visualization is a reconstruction of the surface that was shown to the observers. This can be achieved by applying Eq. (2) for all the pixels of the image (Fig. 3a) and performing triangulation to create a mesh from a point cloud. The same transformation can be applied to a measured set of fixation points $\Phi_e$, which produces the result shown in Fig. 3b. As may be seen, the result meets expectations and looks like traces of human gaze on the object's surface. In order to collate individual points to estimate a spatial distribution, 3D-filtering with a Gaussian kernel can be performed analogously to conventional 2D-filtering of flat maps. In order to simplify computations in continuous space, the points were assigned to a discrete 3D grid of voxels. The grid may have an arbitrary size, which in turn defines the desired resolution of the output map. In MatLab, the obtained values of voxels can be used as colormap indices for faces (or vertices) of the reconstructed surface in order to visualize the density of the distribution via color (Fig. 3c). Another option for visualization is to use voxels' values as a 3D LUT for RGB values of a texture in the rendering engine, in the same way that the maps of coordinates (Fig. 2a) are created.  The resultant colorized meshes are convenient for intuitive visualization of gaze distribution as well as more accurate, in comparison to 2D, spatial analysis of the measured fixations. 

\section{Experimental}
The presented technique of processing of eye-tracking results in 3D was applied for a study of the impact of material appearance on human attention in computer graphics.
\subsection{Stimuli}
Observers were presented with short animations of classical sculptures rotating about their vertical axis (video demonstration: \href{https://drive.google.com/open?id=1nr0nQoQqkrBMGJk9mFh-0UX27G4hyKLC}{[link]}). Rotation of the statues was intended to enhance the perception of shape and material appearance by providing motion parallax, including moving specularities and shadows. 3D models of classical statues were taken from a collection of high-resolution laser-scans created by Oliver Laric\footnote{http://threedscans.com}. Images were created using the Mitsuba physically-based rendering engine, configured for hyperspectral rendering with 31 10-nm wavelength bins between 395 and 705 nm. Then, spectral data was converted to linear RGB using the measured spectral power distribution of the display. The three material conditions "smooth gold", "rough gold", and "imitation gold coating" were configured as follows. For the "smooth gold" condition, the gold Mitsuba material was used with roughness set to zero. For the "rough gold" condition, the ggx roughness distribution was configured to 0.13. For the "imitation gold coating" condition, the silver Mitsuba material was used with zero roughness, set beneath a thin, bumpy coating. The coating was created by scaling up each model by 1\% and applying a displacement modifier configured with Stucco noise. The coating was assigned a refractive index of 1.4, and specified as a "homogeneous medium" with an RGB absorbance of [0,0.10,0.75].\\
Animations were created with each sculpture rotating about its $z$-axis with a sinusoidal motion of a full angle of 50$^{\circ}$ split into 61 frames. The frame rate of the animation was equal to 30 fps. The light probe used is the "Overcast day/building site" environment light probe made by Bernhard Vogel \footnote{http://dativ.at/lightprobes/}. Rendering was carried out with 256 samples using the "Extended volumetric path tracer", to a spatial resolution of 1024 $\times$ 1024 pixels.

\subsection{Experimental setup}
The stimuli were shown to observers on a Cambridge Research Systems (CRS) Display++ in ViSaGe mode located at the distance of 175 cm from the eyes. Fixations were measured using eye-tracker Eyelink1000 (with 2000Hz Extension). \\
The study was approved by the Medical Sciences Interdivisional Research Ethics Committee at the University of Oxford, in accordance with the Declaration of Helsinki. Observers were recruited from the staff and students of the university. There were 12 participants in total (5 males and 7 females), aged between 20-29 years old, all with normal color vision. \\
The animations of sculptures were presented under a set of different tasks (questions):
\begin{enumerate}
  \item Free viewing.
  \item Describe the material the object is made from.
  \item What is the age of the person?
  \item Surmise what the person is thinking.
  \item What clothes the person is wearing?
  \item Remember and describe the pose of the person.
\end{enumerate}

We presented three statues in three different materials. Questions 1-2 are general and were asked for all nine stimuli in random order, whereas questions 3-6 are related to semantic information, thus were asked only once for each statue (using different materials for different observers). The 8-second animations were followed by a response screen that was displayed while participants gave oral answers (not analysed here). 

\section{Results and discussion}

\begin{figure*}[ht!]
  \centering
  \includegraphics[width=0.9\linewidth]{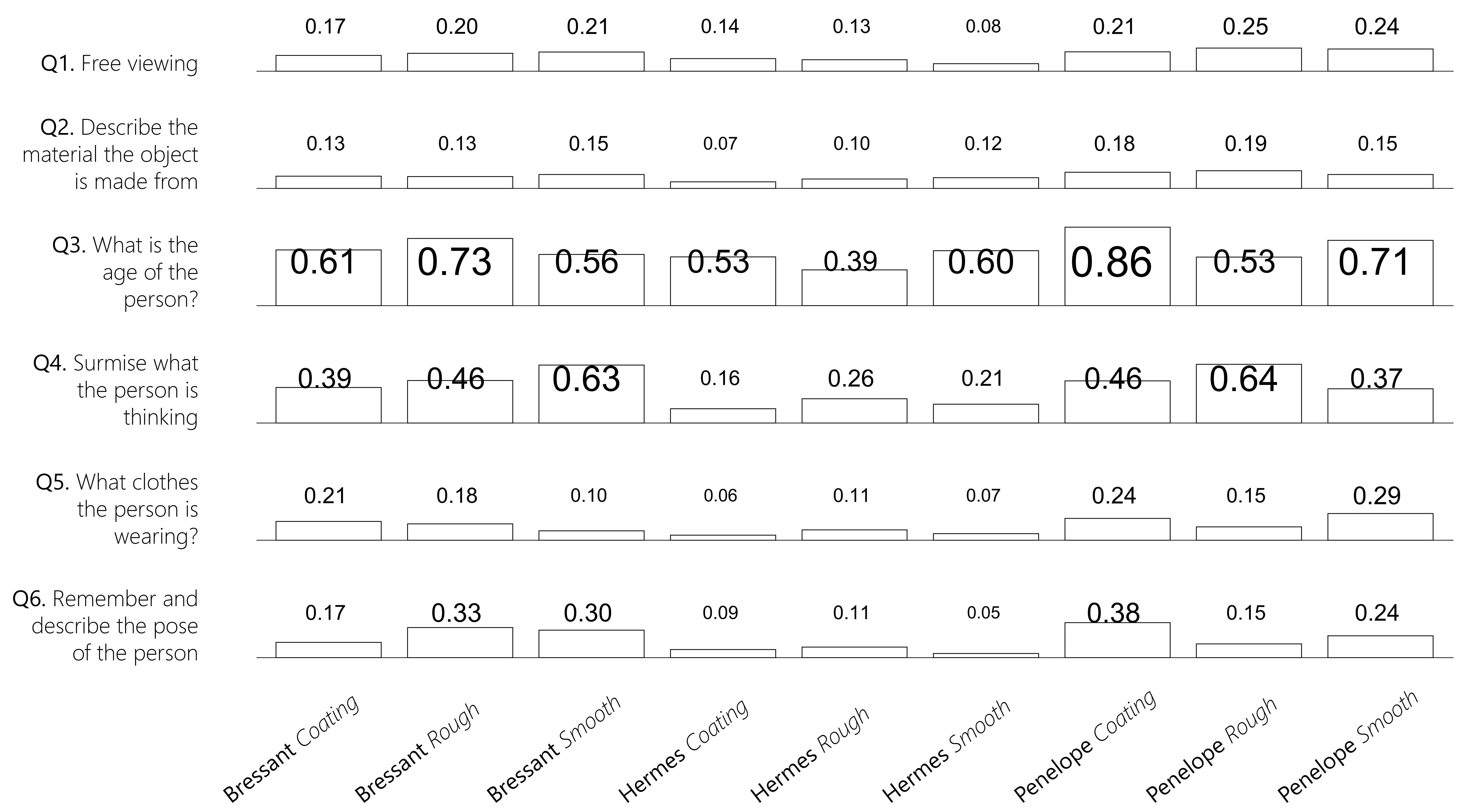}
  \caption{The fraction of gaze-fixations falling on the head of sculpture, while the rest fall on the body. The bar chart in the background visualizes the same values. The fixations of all observers are pooled in one map.}
  \Description{The fraction of gaze-fixations falling on the head of sculpture, while the rest fall on the body. }
  \label{fig:4}
\end{figure*}

\begin{figure*}[t!]
  \centering
  \includegraphics[width=0.9\linewidth]{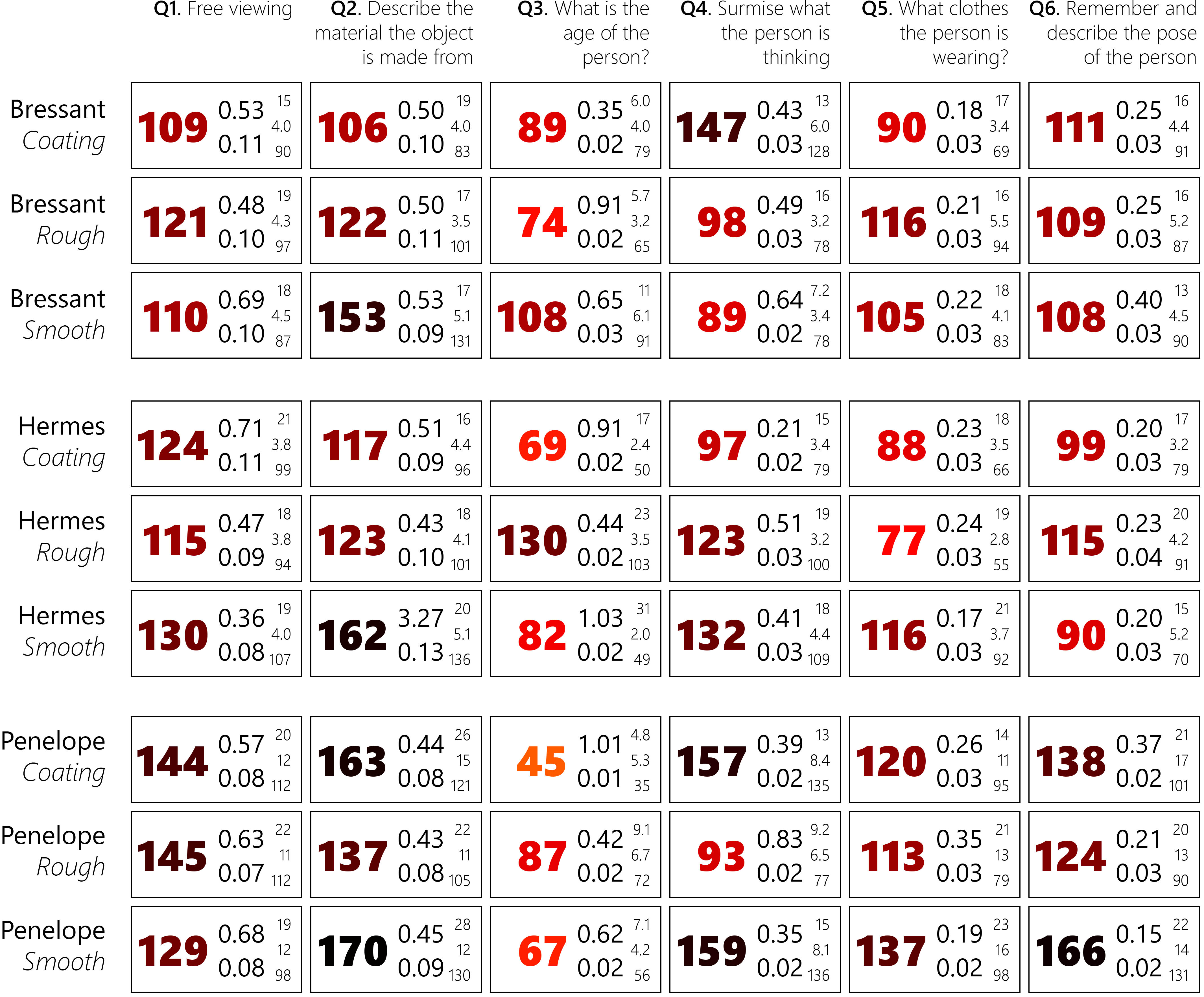}
  \caption{The description of measured distribution of fixations for different combinations of parameters (question, model, material). For each box: left - $\overline{WSS}$; middle - $max(map)$, $mean(map)$; right - $Var(\varphi_x)$, $Var(\varphi_y)$, and $Var(\varphi_z)$.}
  \Description{The description of the measured distribution of fixations for different combinations of parameters.}
  \label{fig:5}
\end{figure*}

According to the design of the experiment, there are multiple factors that influence the result: question, model, and material. Different combinations of these parameters produce 54 ($6\times3\times3$) output maps, which may be accessed online: \href{https://drive.google.com/open?id=1O2EoYUk13Lgnuv451GVnyONQZwHI4VlY}{[link]}. A few of them (question (1), Hermes) are presented in Fig. 1.

\subsection{Impact of a given task}
The classical eye-tracking work of Yarbus demonstrates how the question asked about the painting influences what part of the painting the observers are more likely to look at \cite{yarbus1967eye}. In this section, we want to evaluate how this translates to 3D computer graphics. \\
The reader could notice that the experimental questions (1)-(6) are designed in the following way: questions (1)-(2) are general, questions (3)-(4) stimulate the observer to look at the face of the sculpture, whereas questions (5)-(6) are specifically oriented on the body of the sculpture. Thus, we expect to see a large difference between a number of fixations located on the head and on the body between questions (3)-(4) and (5)-(6). In order to evaluate it quantitatively, we segment the statue into two parts ("head" and "body") and count the total number of fixations from all the observers in each region. Figure 4 presents a table that summarizes our results. We report the fraction of fixations on the "head", while "body" can be found as a complementary part to 1. The bar chart in the background is aimed to facilitate the understanding of the results.\\
As may be seen, the results do agree with the theory of Yarbus. Average "head"-fractions for questions (3) and (4) are 0.613 and 0.400 respectively, whereas the corresponding values for questions (5) and (6) are 0.155 and 0.202. There is no trivial dependence on the model or the material. The only observation is that the Hermes sculpture has fewer fixations on the head than Bressant and Penelope, which may be related to the fact that the head of this sculpture is displaced from the axis of rotation, thus it shifts a significantly larger distance and is harder to follow.

\subsection{Statistics of fixations distribution}
Apart from localization in semantically distinct regions, the obtained maps also differ by the shape of their distribution and density of fixation points. Particularly, some fixations are densely localized, while others are sparse. We characterize these effects by a set of standard metrics. A \textit{compactness} of the distribution may be described by a classical cluster analysis metric Within-cluster Sum of Squares (WSS). Assuming the fixation points belong to one cluster, its centroid $\bar{\varphi_i}$ and mean WSS score can be found as follows:
\begin{equation}
    \overline{WSS} = \frac{1}{N_\varphi}\sum_{i = x,y,z}\sum_{\varphi\in\Phi} \|i_\varphi - \overline{i_\varphi}\|^2 \ \ = \sum_{i = x,y,z}Var(i_\varphi) \ \  ,
\end{equation} 
where $\overline{i_\varphi}=\frac{1}{N_\varphi}\sum_{\varphi\in\Phi}i_\varphi$, and $N_\varphi$ is the total number of fixation points. We also report the variance of each spatial coordinates of the fixations $Var(x_\varphi)$, $Var(y_\varphi)$, and $Var(z_\varphi)$, the sum of which is equal to mean WSS (Eq. 3). After the fixation map is blurred with Gaussian and forms a probability map (saliency map), its values comprehensively describe the density of the fixations distribution. Thus, we also report the maximum value of a distribution (peak density) and the mean value. \\
The results for each combination of experimental parameters are presented in Fig. 5. Each box contains mean WSS (large number, left row), maximum and the mean of the blurred map (top and bottom numbers in the middle column), and variance of x, y, and z coordinates (top to bottom in the right column correspondingly). \\
It may be seen that maps concentrated at the face of a sculpture (Question 3) are the most localized. The difference of the values for rough (matte) and smooth (glossy) materials is not systematic and does not allow general conclusions to be drawn about the impact of material appearance on gaze behaviour.

\subsection{Within-subject comparison}
The previous results were obtained using the fixations of all observers pooled in a single map. This section presents an analysis of the difference in the gaze behaviour of each individual observer with specific regard to the rendered material. Also, to simplify the comparison, we consider only data from question (1) as the most general and natural task. \\
Table 1 presents the scores computed. Each score was computed for a single observer and then averaged between all the observers and all the sculptures. The first score is the difference of compactness of fixations distributions (mean WSS), such that a value of 0 means that both distributions are equally compact. The second and third scores are standard metrics for measuring the similarity of saliency maps: \textit{linear Correlation Coefficient} (CC) \cite{le2007predicting}\cite{Bylinskii2018WhatDD}, 1 or -1 means that maps are correlated, 0 - uncorrelated; and \textit{Similarity} (SIM) also referred to as \textit{histogram
intersection}  \cite{Rubner2000}\cite{riche2013saliency}, such that the value of 1 means that maps are identical, 0 mean that maps are opposite. \\

\begin{table}[b]
  \caption{Pair-wise, within-subjects, comparison of overt attention maps corresponding to different materials.}
  \label{tab:1}
  \begin{tabular}{p{2.9cm}lll}
    \toprule
    & a: Rough & a: Smooth & a: Coating \\
    & b: Smooth & b: Coating & b: Rough \\
    \midrule
    $\overline{WSS}_a - \overline{WSS}_b$ & 11.41 & -6.39 & -2.53 \\
    CC(a,b)                       & 0.348 & 0.328 & 0.349 \\
    SIM(a,b)                & 0.431 & 0.418 & 0.435 \\
  \bottomrule
\end{tabular}
\end{table}

Interestingly, both CC and SIM metrics show almost the same low level of similarity for each of the three pairs, which implies that the fixation distributions on all three materials are equally distinct, and do not correlate with each other. The WSS score shows the largest difference of compactness is between Rough (matte) and Smooth (glossy) materials, thus, we can conclude that using matte renders helps to concentrate the overt attention of the observer, while smooth, shiny surfaces make it more sparsely distributed. The Coating material with high bumpiness produces an intermediate level of compactness. However, these results only describe expected average trends, while particular cases may deviate from it significantly.

\section{Conclusions}
This work presents the development of a new method of re-projection of gaze-fixations measured on a 2D screen to a 3D surface of rendered objects. This significantly increases the accuracy of spatial analysis of fixations, as well as creating intuitively understandable 3D visualizations of the overt attention map. We apply this methodology to studying the relation between gaze behaviour and material appearance in computer graphics. We illustrate that the same 3D models rendered with different parameters of appearance have very different overt attention distributions. Common similarity metrics CC and SIM confirm low correlation between the maps. However, we were not able to detect systematic, generalize patterns of influence of different materials across statues. One of the practical conclusions we could make is that rough matte surfaces concentrate observers' gaze slightly more densely than glossy surfaces, whereas the latter distribute gaze more sparsely. Another is that the task given to observers has a substantial influence on the direction of their gaze, which agrees with findings from conventional 2D eye-tracking. \\
Further development of this work will allow us to collect more gaze data, and to perform more detailed analysis with consideration of other semantic features for the characterization of fixation localization.

\begin{acks}
This work was supported by the Arts and Humanities Research Council grant AH/N001222/1 to HES. OS is funded through the Master Color in Science and Industry (COSI). JSH is funded through the Andrew W. Mellon Foundation and Clarendon Fund.
\end{acks}



\begin{thebibliography}{23}


\ifx \showCODEN    \undefined \def \showCODEN     #1{\unskip}     \fi
\ifx \showDOI      \undefined \def \showDOI       #1{#1}\fi
\ifx \showISBNx    \undefined \def \showISBNx     #1{\unskip}     \fi
\ifx \showISBNxiii \undefined \def \showISBNxiii  #1{\unskip}     \fi
\ifx \showISSN     \undefined \def \showISSN      #1{\unskip}     \fi
\ifx \showLCCN     \undefined \def \showLCCN      #1{\unskip}     \fi
\ifx \shownote     \undefined \def \shownote      #1{#1}          \fi
\ifx \showarticletitle \undefined \def \showarticletitle #1{#1}   \fi
\ifx \showURL      \undefined \def \showURL       {\relax}        \fi
\providecommand\bibfield[2]{#2}
\providecommand\bibinfo[2]{#2}
\providecommand\natexlab[1]{#1}
\providecommand\showeprint[2][]{arXiv:#2}

\bibitem[\protect\citeauthoryear{Anderson}{Anderson}{2011}]%
        {anderson2011visual}
\bibfield{author}{\bibinfo{person}{Barton~L Anderson}.}
  \bibinfo{year}{2011}\natexlab{}.
\newblock \showarticletitle{Visual perception of materials and surfaces}.
\newblock \bibinfo{journal}{\emph{Current Biology}} \bibinfo{volume}{21},
  \bibinfo{number}{24} (\bibinfo{year}{2011}), \bibinfo{pages}{R978--R983}.
\newblock


\bibitem[\protect\citeauthoryear{Bergstrom and Schall}{Bergstrom and
  Schall}{2014}]%
        {bergstrom2014eye}
\bibfield{author}{\bibinfo{person}{Jennifer~Romano Bergstrom} {and}
  \bibinfo{person}{Andrew Schall}.} \bibinfo{year}{2014}\natexlab{}.
\newblock \bibinfo{booktitle}{\emph{Eye tracking in user experience design}}.
\newblock \bibinfo{publisher}{Elsevier}.
\newblock


\bibitem[\protect\citeauthoryear{Bylinskii, Judd, Oliva, Torralba, and
  Durand}{Bylinskii et~al\mbox{.}}{2018}]%
        {Bylinskii2018WhatDD}
\bibfield{author}{\bibinfo{person}{Zoya Bylinskii}, \bibinfo{person}{Tilke
  Judd}, \bibinfo{person}{Aude Oliva}, \bibinfo{person}{Antonio Torralba},
  {and} \bibinfo{person}{Fr{\'e}do Durand}.} \bibinfo{year}{2018}\natexlab{}.
\newblock \showarticletitle{What Do Different Evaluation Metrics Tell Us About
  Saliency Models?}
\newblock \bibinfo{journal}{\emph{IEEE Transactions on Pattern Analysis and
  Machine Intelligence}}  \bibinfo{volume}{41} (\bibinfo{year}{2018}),
  \bibinfo{pages}{740--757}.
\newblock


\bibitem[\protect\citeauthoryear{Chadwick and Kentridge}{Chadwick and
  Kentridge}{2015}]%
        {chadwick2015perception}
\bibfield{author}{\bibinfo{person}{AC Chadwick} {and} \bibinfo{person}{RW
  Kentridge}.} \bibinfo{year}{2015}\natexlab{}.
\newblock \showarticletitle{The perception of gloss: A review}.
\newblock \bibinfo{journal}{\emph{Vision research}}  \bibinfo{volume}{109}
  (\bibinfo{year}{2015}), \bibinfo{pages}{221--235}.
\newblock


\bibitem[\protect\citeauthoryear{Cowen, Ball, and Delin}{Cowen
  et~al\mbox{.}}{2002}]%
        {cowen2002eye}
\bibfield{author}{\bibinfo{person}{Laura Cowen}, \bibinfo{person}{Linden~Js
  Ball}, {and} \bibinfo{person}{Judy Delin}.} \bibinfo{year}{2002}\natexlab{}.
\newblock \showarticletitle{An eye movement analysis of web page usability}.
\newblock In \bibinfo{booktitle}{\emph{People and computers XVI-memorable yet
  invisible}}. \bibinfo{publisher}{Springer}, \bibinfo{pages}{317--335}.
\newblock


\bibitem[\protect\citeauthoryear{Fleming}{Fleming}{2014}]%
        {fleming2014visual}
\bibfield{author}{\bibinfo{person}{Roland~W Fleming}.}
  \bibinfo{year}{2014}\natexlab{}.
\newblock \showarticletitle{Visual perception of materials and their
  properties}.
\newblock \bibinfo{journal}{\emph{Vision research}}  \bibinfo{volume}{94}
  (\bibinfo{year}{2014}), \bibinfo{pages}{62--75}.
\newblock


\bibitem[\protect\citeauthoryear{Howlett, Hamill, and O'Sullivan}{Howlett
  et~al\mbox{.}}{2005}]%
        {howlett2005predicting}
\bibfield{author}{\bibinfo{person}{Sarah Howlett}, \bibinfo{person}{John
  Hamill}, {and} \bibinfo{person}{Carol O'Sullivan}.}
  \bibinfo{year}{2005}\natexlab{}.
\newblock \showarticletitle{Predicting and evaluating saliency for simplified
  polygonal models}.
\newblock \bibinfo{journal}{\emph{ACM Transactions on Applied Perception
  (TAP)}} \bibinfo{volume}{2}, \bibinfo{number}{3} (\bibinfo{year}{2005}),
  \bibinfo{pages}{286--308}.
\newblock


\bibitem[\protect\citeauthoryear{Jacob and Karn}{Jacob and Karn}{2003}]%
        {Jacob2003-JACETI}
\bibfield{author}{\bibinfo{person}{Robert J.~K. Jacob} {and}
  \bibinfo{person}{Keith~S. Karn}.} \bibinfo{year}{2003}\natexlab{}.
\newblock \showarticletitle{Eye Tracking in Human-Computer Interaction and
  Usability Research: Ready to Deliver the Promises}.
\newblock \bibinfo{journal}{\emph{Mind}} \bibinfo{volume}{2},
  \bibinfo{number}{3} (\bibinfo{year}{2003}), \bibinfo{pages}{4}.
\newblock


\bibitem[\protect\citeauthoryear{Kentridge, Thomson, Heywood,
  et~al\mbox{.}}{Kentridge et~al\mbox{.}}{2012}]%
        {kentridge2012glossiness}
\bibfield{author}{\bibinfo{person}{Robert~W Kentridge},
  \bibinfo{person}{Rebecca Thomson}, \bibinfo{person}{Charles~A Heywood},
  {et~al\mbox{.}}} \bibinfo{year}{2012}\natexlab{}.
\newblock \showarticletitle{Glossiness perception can be mediated independently
  of cortical processing of colour or texture}.
\newblock \bibinfo{journal}{\emph{Cortex}} \bibinfo{volume}{48},
  \bibinfo{number}{9} (\bibinfo{year}{2012}), \bibinfo{pages}{1244--1246}.
\newblock


\bibitem[\protect\citeauthoryear{Kim, Varshney, Jacobs, and
  Guimbreti{\`e}re}{Kim et~al\mbox{.}}{2010}]%
        {kim2010mesh}
\bibfield{author}{\bibinfo{person}{Youngmin Kim}, \bibinfo{person}{Amitabh
  Varshney}, \bibinfo{person}{David~W Jacobs}, {and}
  \bibinfo{person}{Fran{\c{c}}ois Guimbreti{\`e}re}.}
  \bibinfo{year}{2010}\natexlab{}.
\newblock \showarticletitle{Mesh saliency and human eye fixations}.
\newblock \bibinfo{journal}{\emph{ACM Transactions on Applied Perception
  (TAP)}} \bibinfo{volume}{7}, \bibinfo{number}{2} (\bibinfo{year}{2010}),
  \bibinfo{pages}{12}.
\newblock


\bibitem[\protect\citeauthoryear{Landy}{Landy}{2007}]%
        {landy2007visual}
\bibfield{author}{\bibinfo{person}{Michael~S Landy}.}
  \bibinfo{year}{2007}\natexlab{}.
\newblock \showarticletitle{Visual perception: A gloss on surface properties}.
\newblock \bibinfo{journal}{\emph{Nature}} \bibinfo{volume}{447},
  \bibinfo{number}{7141} (\bibinfo{year}{2007}), \bibinfo{pages}{158}.
\newblock


\bibitem[\protect\citeauthoryear{Lavou{\'e}, Cordier, Seo, and
  Larabi}{Lavou{\'e} et~al\mbox{.}}{2018}]%
        {lavoue2018visual}
\bibfield{author}{\bibinfo{person}{Guillaume Lavou{\'e}},
  \bibinfo{person}{Fr{\'e}d{\'e}ric Cordier}, \bibinfo{person}{Hyewon Seo},
  {and} \bibinfo{person}{Mohamed-Chaker Larabi}.}
  \bibinfo{year}{2018}\natexlab{}.
\newblock \showarticletitle{Visual attention for rendered 3D shapes}. In
  \bibinfo{booktitle}{\emph{Computer Graphics Forum}},
  Vol.~\bibinfo{volume}{37}. Wiley Online Library, \bibinfo{pages}{191--203}.
\newblock


\bibitem[\protect\citeauthoryear{Le~Meur, Le~Callet, and Barba}{Le~Meur
  et~al\mbox{.}}{2007}]%
        {le2007predicting}
\bibfield{author}{\bibinfo{person}{Olivier Le~Meur}, \bibinfo{person}{Patrick
  Le~Callet}, {and} \bibinfo{person}{Dominique Barba}.}
  \bibinfo{year}{2007}\natexlab{}.
\newblock \showarticletitle{Predicting visual fixations on video based on
  low-level visual features}.
\newblock \bibinfo{journal}{\emph{Vision research}} \bibinfo{volume}{47},
  \bibinfo{number}{19} (\bibinfo{year}{2007}), \bibinfo{pages}{2483--2498}.
\newblock


\bibitem[\protect\citeauthoryear{Leek, Cristino, Conlan, Patterson, Rodriguez,
  and Johnston}{Leek et~al\mbox{.}}{2012}]%
        {10.1167/12.1.7}
\bibfield{author}{\bibinfo{person}{E.~Charles Leek}, \bibinfo{person}{Filipe
  Cristino}, \bibinfo{person}{Lina~I. Conlan}, \bibinfo{person}{Candy
  Patterson}, \bibinfo{person}{Elly Rodriguez}, {and}
  \bibinfo{person}{Stephen~J. Johnston}.} \bibinfo{year}{2012}\natexlab{}.
\newblock \showarticletitle{{Eye movement patterns during the recognition of
  three-dimensional objects: Preferential fixation of concave surface curvature
  minima}}.
\newblock \bibinfo{journal}{\emph{Journal of Vision}} \bibinfo{volume}{12},
  \bibinfo{number}{1} (\bibinfo{date}{01} \bibinfo{year}{2012}),
  \bibinfo{pages}{7--7}.
\newblock
\showISSN{1534-7362}
\urldef\tempurl%
\url{https://doi.org/10.1167/12.1.7}
\showDOI{\tempurl}
\showeprint{https://jov.arvojournals.org/arvo/content\_public/journal/jov/933488/jov-12-1-7.pdf}


\bibitem[\protect\citeauthoryear{Leonards, Baddeley, Gilchrist, Troscianko,
  Ledda, and Williamson}{Leonards et~al\mbox{.}}{2007}]%
        {leonards2007mediaeval}
\bibfield{author}{\bibinfo{person}{Ute Leonards}, \bibinfo{person}{Roland
  Baddeley}, \bibinfo{person}{Iain~D Gilchrist}, \bibinfo{person}{Tom
  Troscianko}, \bibinfo{person}{Patrick Ledda}, {and} \bibinfo{person}{Beth
  Williamson}.} \bibinfo{year}{2007}\natexlab{}.
\newblock \showarticletitle{Mediaeval artists: Masters in directing the
  observers' gaze}.
\newblock \bibinfo{journal}{\emph{Current Biology}} \bibinfo{volume}{17},
  \bibinfo{number}{1} (\bibinfo{year}{2007}), \bibinfo{pages}{R8--R9}.
\newblock


\bibitem[\protect\citeauthoryear{Mantiuk, Bazyluk, and Mantiuk}{Mantiuk
  et~al\mbox{.}}{2013}]%
        {mantiuk2013gaze}
\bibfield{author}{\bibinfo{person}{Radoslaw Mantiuk}, \bibinfo{person}{Bartosz
  Bazyluk}, {and} \bibinfo{person}{Rafal~K Mantiuk}.}
  \bibinfo{year}{2013}\natexlab{}.
\newblock \showarticletitle{Gaze-driven Object Tracking for Real Time
  Rendering}. In \bibinfo{booktitle}{\emph{Computer Graphics Forum}},
  Vol.~\bibinfo{volume}{32}. Wiley Online Library, \bibinfo{pages}{163--173}.
\newblock


\bibitem[\protect\citeauthoryear{Qi, Xu, Shang, and Dong}{Qi
  et~al\mbox{.}}{2018}]%
        {qi2018fusing}
\bibfield{author}{\bibinfo{person}{Lin Qi}, \bibinfo{person}{Ying Xu},
  \bibinfo{person}{Xiaowei Shang}, {and} \bibinfo{person}{Junyu Dong}.}
  \bibinfo{year}{2018}\natexlab{}.
\newblock \showarticletitle{Fusing Visual Saliency for Material Recognition}.
  In \bibinfo{booktitle}{\emph{Proceedings of the IEEE Conference on Computer
  Vision and Pattern Recognition Workshops}}. \bibinfo{pages}{1965--1968}.
\newblock


\bibitem[\protect\citeauthoryear{Rayner, Rotello, Stewart, Keir, and
  Duffy}{Rayner et~al\mbox{.}}{2001}]%
        {rayner2001integrating}
\bibfield{author}{\bibinfo{person}{Keith Rayner}, \bibinfo{person}{Caren~M
  Rotello}, \bibinfo{person}{Andrew~J Stewart}, \bibinfo{person}{Jessica Keir},
  {and} \bibinfo{person}{Susan~A Duffy}.} \bibinfo{year}{2001}\natexlab{}.
\newblock \showarticletitle{Integrating text and pictorial information: eye
  movements when looking at print advertisements.}
\newblock \bibinfo{journal}{\emph{Journal of experimental psychology: Applied}}
  \bibinfo{volume}{7}, \bibinfo{number}{3} (\bibinfo{year}{2001}),
  \bibinfo{pages}{219}.
\newblock


\bibitem[\protect\citeauthoryear{Riche, Duvinage, Mancas, Gosselin, and
  Dutoit}{Riche et~al\mbox{.}}{2013}]%
        {riche2013saliency}
\bibfield{author}{\bibinfo{person}{Nicolas Riche}, \bibinfo{person}{Matthieu
  Duvinage}, \bibinfo{person}{Matei Mancas}, \bibinfo{person}{Bernard
  Gosselin}, {and} \bibinfo{person}{Thierry Dutoit}.}
  \bibinfo{year}{2013}\natexlab{}.
\newblock \showarticletitle{Saliency and human fixations: State-of-the-art and
  study of comparison metrics}. In \bibinfo{booktitle}{\emph{Proceedings of the
  IEEE international conference on computer vision}}.
  \bibinfo{pages}{1153--1160}.
\newblock


\bibitem[\protect\citeauthoryear{Rubner, Tomasi, and Guibas}{Rubner
  et~al\mbox{.}}{2000}]%
        {Rubner2000}
\bibfield{author}{\bibinfo{person}{Yossi Rubner}, \bibinfo{person}{Carlo
  Tomasi}, {and} \bibinfo{person}{Leonidas~J. Guibas}.}
  \bibinfo{year}{2000}\natexlab{}.
\newblock \showarticletitle{The Earth Mover's Distance as a Metric for Image
  Retrieval}.
\newblock \bibinfo{journal}{\emph{International Journal of Computer Vision}}
  \bibinfo{volume}{40}, \bibinfo{number}{2} (\bibinfo{date}{01 Nov}
  \bibinfo{year}{2000}), \bibinfo{pages}{99--121}.
\newblock
\showISSN{1573-1405}
\urldef\tempurl%
\url{https://doi.org/10.1023/A:1026543900054}
\showDOI{\tempurl}


\bibitem[\protect\citeauthoryear{Wada, Sakano, and Ando}{Wada
  et~al\mbox{.}}{2014}]%
        {wada2014human}
\bibfield{author}{\bibinfo{person}{Atsushi Wada}, \bibinfo{person}{Yuichi
  Sakano}, {and} \bibinfo{person}{Hiroshi Ando}.}
  \bibinfo{year}{2014}\natexlab{}.
\newblock \showarticletitle{Human cortical areas involved in perception of
  surface glossiness}.
\newblock \bibinfo{journal}{\emph{NeuroImage}}  \bibinfo{volume}{98}
  (\bibinfo{year}{2014}), \bibinfo{pages}{243--257}.
\newblock


\bibitem[\protect\citeauthoryear{Wang, Koch, Holmqvist, and Alexa}{Wang
  et~al\mbox{.}}{2018}]%
        {wang2018tracking}
\bibfield{author}{\bibinfo{person}{Xi Wang}, \bibinfo{person}{Sebastian Koch},
  \bibinfo{person}{Kenneth Holmqvist}, {and} \bibinfo{person}{Marc Alexa}.}
  \bibinfo{year}{2018}\natexlab{}.
\newblock \showarticletitle{Tracking the gaze on objects in 3D: how do people
  really look at the bunny?}. In \bibinfo{booktitle}{\emph{SIGGRAPH Asia 2018
  Technical Papers}}. ACM, \bibinfo{pages}{188}.
\newblock


\bibitem[\protect\citeauthoryear{Yarbus}{Yarbus}{1967}]%
        {yarbus1967eye}
\bibfield{author}{\bibinfo{person}{Alfred~L Yarbus}.}
  \bibinfo{year}{1967}\natexlab{}.
\newblock \bibinfo{booktitle}{\emph{Eye movements and vision}}.
\newblock \bibinfo{publisher}{Plenum}, \bibinfo{address}{New York, NY, USA}.
\newblock


\end{thebibliography}
\end{document}